%% file: main.tex
\documentclass[graybox]{svmult}

\usepackage{type1cm}        
\usepackage{makeidx}         
\usepackage{graphicx}        
\usepackage{multicol}        
\usepackage[bottom]{footmisc}

\usepackage{newtxtext}       %
\usepackage{newtxmath}       

\makeindex             

\begin{document}

\title*{Game Theory Based Privacy Preserving Approach for Collaborative Deep Learning in IoT}

\titlerunning{Game Theory Based Privacy Preserving Approach for CDL in IoT}

\author{Deepti Gupta, Smriti Bhatt, Paras Bhatt, Maanak Gupta, and Ali Saman Tosun}

\authorrunning{Deepti Gupta et al.}
\institute{Deepti Gupta \and Ali Saman Tosun\at Department of Computer Science, University of Texas at San Antonio, San Antonio, Texas 78249, USA , \email{deepti.mrt@gmail.com; ali.tosun@utsa.edu}
\and Smriti Bhatt \at Department of Computing and Cyber Security, Texas A \& M University-San Antonio, San Antonio, Texas 78224, USA. \email{sbhatt@tamusa.edu}
\and Paras Bhatt \at Department of Information Systems and Cyber Security, University of Texas at San Antonio, San Antonio, Texas 78249, USA. \email{paras.bhatt@utsa.edu}
\and Maanak Gupta \at Department of Computer Science, Tennessee Technological University, Cookeville, Tennessee 38505, USA. \email{mgupta@tntech.edu}}
%
%
\maketitle

\abstract{The exponential growth of Internet of Things (IoT) has become a transcending force in creating innovative smart devices and connected domains including smart homes, healthcare, transportation and manufacturing. With billions of IoT devices, there is a huge amount of data continuously being generated, transmitted, and stored at various points in the IoT architecture. Deep learning is widely being used in IoT applications to extract useful insights from IoT data. However, IoT users have security and privacy concerns and prefer not to share their personal data with third party applications or stakeholders. In order to address user privacy concerns, Collaborative Deep Learning (CDL) has been largely employed in data-driven applications which enables multiple IoT devices to train their models locally on edge gateways.
In this chapter, we first discuss different types of deep learning approaches and how these approaches can be employed in the IoT domain. We present a privacy-preserving collaborative deep learning approach for IoT devices which can achieve benefits from other devices in the system. This learning approach is analyzed from the behavioral perspective of mobile edge devices using a game-theoretic model. 
We analyze the Nash Equilibrium in \textit{N}-player static game model. We further present a novel fair collaboration strategy among edge IoT devices using cluster based approach to solve the CDL game, which enforces mobile edge devices for cooperation. We also present implementation details and evaluation analysis in a real-world smart home deployment.}
\input{introduction}
\input{Literature_Review}

\input{Deep_Learning}
\input{System_Model}

\input{Deep_Learning_Game}
\input{Implementation}
\input{Conclusion}
    {
    \bibliographystyle{plain}
    \bibliography{reference}
    }

\end{document}

%% file: introduction.tex
\section{Introduction}
\label{introduction}
In recent years, Internet of Things (IoT) is becoming a pervasive reality of our lives with billions of IoT devices which are continuously increasing in number. These smart devices are connected to or associated with users and generate huge amount of data, from user health information \cite{baker2017internet} to social networking \cite{atzori2012social} information of the users. This large amount of valuable data enables to utilize Deep Learning (DL) models for training and enhancing intelligence of various data-driven IoT applications. Generally, these devices are resource constraint and leverage the cloud computing services and platform for expanding their storage and analytics power \cite{bhatt2017access1,bhatt2019authorizations}. Thus, most of the IoT devices connect to a central cloud platform \cite{gupta2020access,gupta2019dynamic,gupta2020attribute,awaysheh2020next} to use remote services. These services are crucial for storage of the datasets and learning Machine Learning (ML) models. However, there is additional latency incurred while these smart devices interact with the cloud services. To overcome this issue, edge computing is shaping a new paradigm enabling low latency real-time communications between IoT devices and edge devices, such as gateways (e.g., smartphones). These edge devices, also known as edge cloudlets \cite{gupta2020secure,bhatt2017access}, will then communicate with the cloud services. Edge devices are performed data training locally and are also be employed to preserve privacy of personal data. Unlike constrained IoT devices, these edge gateways have the capability to support ML models. A simple example would be a video enabled doorbell that performs training on its local datasets, and identifies person at the door. 

DL models are often associated with the size of training dataset. While training a learning mechanism or model, large training data will enhance the accuracy and performance of a trained model. In today's connected world and new era of big data, data is often distributed across several smart devices, edge cloudlets, and cloud and it cannot be brought together due to user privacy constraints. Collaborative Deep Learning (CDL) allows multiple IoT devices to train their models, without exposing any associated sensitive and private data. CDL offers an attractive trade-off between user privacy and utility of data sets.

Recent research \cite{jiang2019lightweight,chen2019communication,gupta2020security,gupta2021future} have discussed the privacy issues of local training devices and the impact of communication latency between edge gateways and Parameter Server (PS). However, the strategic behavior of the rational local training gateways have not been discussed in previous research, i.e., the authors have assumed that all IoT devices are altruistic. Altruistic devices are ones which always follow a suggested protocol as decided initially, regardless of whether they are benefiting or not by following the specific protocol. However, in a real-world scenario, devices are not altruistic, they are rational. Rational devices are the ones which will deviate from suggested protocol if they think that they will be benefited more by following a different protocol. In our proposed system model, we assume that all mobile edge devices or edge gateways are rational.

Generally, a mobile edge device that has a low quality data, always wants to be a part of CDL to increase accuracy of their local model. Whereas other mobile edge devices have high quality data, do not want to collaborate with low quality data holder mobile edge devices due to privacy concerns while sharing their local gradients. Therefore, there is a dilemma for mobile edge devices to be part or not of CDL. In this chapter, we address the problem of learner's dilemma by proposing a CDL game model and a novel fair collaboration strategy which enables each participant to cooperate in CDL based on the clusters formed to achieve overall benefit to itself in training the local ML model. We also evaluate our CDL game model and novel fair collaboration strategy in smart home deployment using ARAS dataset \cite{alemdar2013aras}. The main contributions of this work are summarized as follows. 
\begin{enumerate}
    \item We identify the problem of unfair cooperation of participants in CDL. In other words, a local training device, which has low quality data and builds its learning model to take advantage from other devices which has high quality data.
    \item We propose a game-theoretic model for analyzing the behavior of mobile edge IoT devices, where each device aims at maximizing the accuracy of its local model with minimal cost of participation in CDL.
    \item We propose a novel fair collaboration strategy for addressing the issue of unfair cooperation in CDL between rational IoT devices.  
    \item We also implement our fair collaboration strategy on ARAS dataset \cite{alemdar2013aras}, and the results reflect that proposed solution elicit cooperation in CDL.
\end{enumerate}
The rest of the chapter is organized as follows. Section \ref{sec:related}
presents relevant work and related background. 
Section \ref{sec:DeepIoT} presents different types of DL techniques applicable in IoT. The System model along with rational assumptions are discussed in Section \ref{sec:model}. Game model and game analysis are explained in Section \ref{sec:game}. Section \ref{sec:analysis} presents implementation of proposed system model on ARAS smart home data along with the analysis of results. Section \ref{sec:conclusion} concludes this work with future research directions. 

%% file: Literature_Review.tex
\section{Background and Related Work}
\label{sec:related}
In this section, we describe related work on information leakage on deep learning models in IoT, and give a brief overview of privacy preserving techniques. Here, we also discuss game theoretical models, which have been used to secure user's personal data.

\subsection{Information Leakage on Deep Learning Models in IoT}

Information leakage of user's personal data has become a well known problem for deep learning models. It is a common problem of accidentally revealing the personal information of individuals. To avoid this, various data masking techniques such as \textit{pseudonymize} and \textit{anonymize} have been user to secure the data towards ensuring user privacy. It is critical to understand the difference between pseudonymized data and anonymized data. In pseudonymize, there is possibility to trace back data into its original state, whereas it becomes impossible to get back data into its original form in anonymize. However, data can trace back data into its original form indirectly. For instance, Netflix released a hundred million anonymized film ratings that included a unique subscriber ID, the movie title, year of release and the date on which the subscriber rated the movie. This anonymized Netflix dataset was matched with data crawled from the Internet Movie Database (IMDb). Even with a small sample of 50 IMDb users it was easy to identify the records of two users. Hence, pseudonymization and anonymization approaches are still vulnerable to some inference attacks that would compromise user data privacy. 

Today, big Internet giants including Google and Amazon are already offering \textit{Machine Learning as a service} and any customer with a specific dataset and a data classification task can upload this dataset to the service and pay it to construct a ML model. This model is then made available to the customer, typically as a black-box API, which is vulnerable to attacks where the adversary can observe the model prediction; however, they cannot access the model parameters, nor any computation. But there are still some possibility of data leakage at cloud platform. The membership inference attack on black-box API is discussed in \cite{shokri2017membership,yeom2018privacy}. The attacker queries the target model with a data record and obtains the model’s prediction on that record. An adversary can also build an algorithm to trace the model’s training dataset of data holders. Rahman et al. \cite{rahman2018membership} show that differential private deep model could also fail against membership inference attack.
Similarly, a novel white-box membership inference attack was proposed by Nasr et al. \cite{nasr2018comprehensive}. This attack measures their training datasets membership leakage against deep learning algorithms. In the white-box attack, the adversary has access to the full model including model prediction, model parameters, and intermediate computation at all different layers. Melis et al. \cite{melis2019exploiting} demonstrate that the updated parameter leaks unintended information about data holders' training datasets; thus, develops passive and active inference attacks to exploit this leakage.

\subsection{Privacy Preserving Deep Learning}
In a collaborative model, each participant has its own sensitive datasets and various privacy mechanisms have been proposed to preserve privacy and protect against exchanging parameters such as Secure Multi-party Communication (SMC) \cite{kerschbaum2009practical}, Homomorphic Encryption (HE) \cite{rivest1978data}, and Differential Privacy \cite{dwork2014algorithmic}. SMC helps to protect intermediate steps of the computation when multiple parties perform collaborative ML on their proprietary inputs. Mohassel et al. \cite{mohassel2017secureml} adopt a two-server model for privacy-preserving training, commonly used by previous work on privacy-preserving deep learning via SMC \cite{gascon2016secure,nikolaenko2013privacy1, nikolaenko2013privacy2}. In this model, during the setup phase, the data holders process and encrypt their data among two non-colluding servers, and during the computation phase, the two servers can train various models on the data holders' joint data without learning any information beyond the trained model. However, Aono et al. \cite{aono2018privacy,aono2017privacy} showed that the local data information may be actually leaked to an honest-but-curious server. To obscure an individual's identity, Differential Privacy (DP) adds mathematical noise to a small sample of the individual's usage pattern. Prior work \cite{abadi2016deep,jiang2019lightweight,shokri2015privacy,weng2018deepchain} have employed DP on privacy-preserving collaborative deep learning system to protect privacy of training data. However, Hitaj et al. \cite{hitaj2017deep} pointed out that the above mentioned work failed to protect data privacy and demonstrated that a curious parameter server can learn private data through Generative Adversarial Networks (GAN) learning.

With the deep learning approaches, a dominant technique to optimize the loss function is Stochastic Gradient Descent (SGD). SGD is a method to find the optimal parameter configuration for a ML algorithm. It iteratively makes small adjustments to a ML network configuration to decrease the error of the network. It has been applied in various privacy-preserving DL models in the literature \cite{abadi2016deep,melis2019exploiting, mohassel2017secureml,nasr2018comprehensive}. Moreover, a distributed selective SGD \cite{shokri2015privacy} assumes two or more data holders training independently and concurrently. After each round of local training, data holders share their gradients asynchronously. On the other hand, downpour SGD \cite{aono2017privacy} is a variant of asynchronous SGD, where a global vector for neural network is initialized randomly. At each iteration, replicas of neural networks are run over local datasets, and the corresponding local gradient vector is sent to the server. A subset or gradient of the local model is shared with a server. The server receives the gradients from data holders by using different approaches like round robin, random order \cite{shokri2015privacy}, cosine distance \cite{chen2018machine}, time based \cite{weng2018deepchain}. The server then aggregates these received parameters using FederatedAveraging algorithm \cite{mcmahan2016communication}, and weighted aggregation strategy \cite{chen2018machine}. One of the other challenges in CDL is to reduce the client-server communication. A temporally weighted aggregation strategy is introduced for less communication cost and high model accuracy. While a number of privacy-preserving solutions exist for collaborating organizations to securely aggregate the parameters in the process of training the models, a rational framework for the participants is discussed in \cite{wu2017game} that balances privacy loss and accuracy gain in their collaboration.

While most of the prior research focus on designing the optimal privacy mechanisms, there is a major requirement to choose a particular privacy mechanism for a particular dataset, such as non-IID dataset, high-quality dataset, and low-quality dataset. 
Zhao et al. \cite{zhao2018privacy} proposed a solution to reduce the impact of low quality data holders. Although there have been a number of privacy mechanisms focusing on data protection, these studies are limited to specific scenarios, for example privacy of exchanging parameters, which makes it difficult to apply these techniques to protect and ensure privacy for the whole training dataset.

\subsection{Game Theory}
Game theory has been applied to data privacy game for analyzing privacy and accuracy. Pejo et al. \cite{pejo2018price} defined two player game, where one player is privacy concerned and other player is not. Esposito  et al. \cite{esposito2018securing} proposed a game model to analyze the interaction between a provider (global ML model) and a requester (local ML model) within a CDL model. Gupta et al. \cite{gupta2020learner, gupta2020game} presented CDL for rational players in their prior work. In this chapter, we construct a game model for rational mobile edge devices cooperation in CDL and present an analysis of the game.



%% file: Deep_Learning.tex
\section{Deep Learning in IoT}
\label{sec:DeepIoT}
IoT architecture, by design, is associated with the generation of multifaceted data. From sensors to automated responses, IoT devices output a swarm of data points that are particularly well suited for ML tasks. Within the domain of ML, the advancing field of DL has of late started using a variety of techniques for harnessing the power of these millions of data points available from IoT sensors and devices. With reference to DL, we discuss some of the most popular techniques as follow.

\begin{figure}[t]
\centering
\includegraphics[width=1\textwidth, height=.5\textheight]{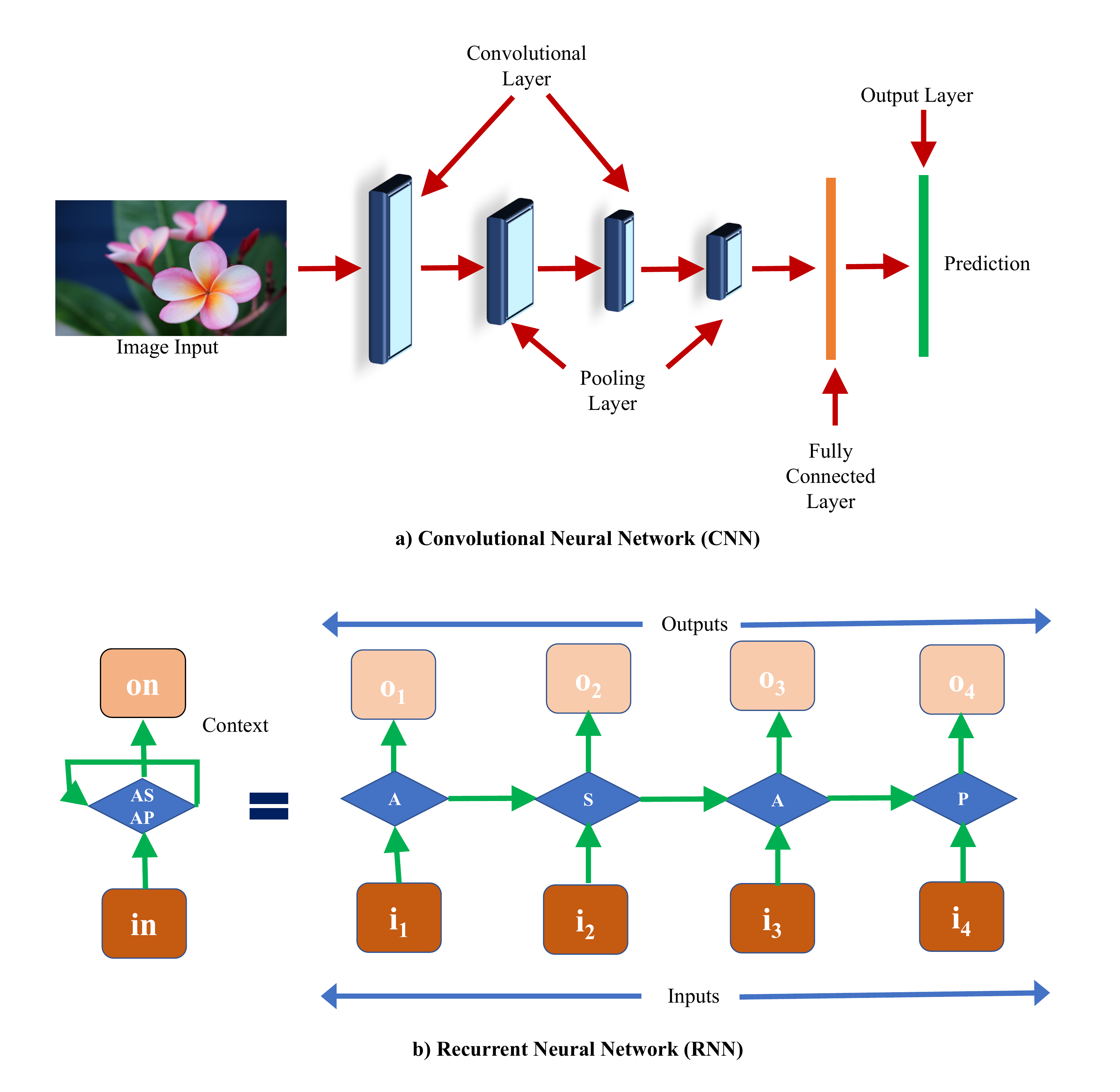}
\caption{A Diagrammatic View of CNN and RNN}
\label{fig:CNNRNN}
\end{figure}

\subsection{Convolutional Neural Network (CNN)}
In its basic form, a CNN DL algorithm differentiates among a set of input images to output a desired category of images as specified by a user. A CNN model derives its efficiency from the ability to learn filters and features in data, that would otherwise have to be hand engineered. With enough training, a CNN model can do that on its own. Using hand engineered features by a programmer to curate learning that is aimed at solving a specific problem cannot to be generalized to solve other similar problems. A powerful general-purpose learning procedure, that replaces the programmer with a set of predefined algorithms which can then be used to obtain effective solutions for a much wider problem domain. This is where the true power of CNN lies. With sufficient data and computing power, learning trumps programming by a long shot \cite{krizhevsky2017imagenet}.

Figure \ref{fig:CNNRNN} a) represents a generic view of the layers in a CNN model which correspond to - 1) Convolutional layer which is responsible for extracting high level features from a set of input images; 2) Pooling Layer which is aimed at reducing the computational power needed for processing the data; and, 3) Fully Connected Layer that is appended for classifying the desired set of images with the help of a softmax function.
Recently, CNN has been widely used in the IoT domain. A classic example is of drones which have cameras mounted that can collect images of crops, traffic on road, or even land use. These images can then be collated to form a large dataset that can be then used to predict crop diseases, traffic congestion, or drought conditions as per specific application domains. It has become a popular deep learning tool with its various instantiations being used across different IoT domains including precision agriculture \cite{milioto2018real}, smart traffic management \cite{pan2017spatial}, and medical diagnosis \cite{liu2017tx}.

\subsection{Recurrent Neural Network (RNN)}

A Recurrent Neural Network (RNN) can be understood as a learning model that has an internal memory. As the name suggests, an RNN model is recurrent by design where it performs the same function for every input. Here, whilst producing an output, the model maintains a copy of that output in the internal memory and then during prediction it uses both the current input as well as the past output to make a decision. There are certain variants of this model which expand on this internal memory concepts primarily Long Short Term Memory (LSTM) models \cite{kayode2020towards}.
Figure \ref{fig:CNNRNN} b) presents a RNN model architecture. In the first step as shown in the figure, the model receives an initial input, then it produces a corresponding output based on it. Then subsequently for the next step the model uses the output obtained in the first step plus a new input for the next step. In this manner the model keeps in mind the context of the training process.

RNN has been applied in several IoT domains including IoT security. More specifically, it has been used to detect attacks against IoT-connected smart home environments \cite{farsi2020application}. In the context of Industrial IoT, RNN has been used to predict maintenance needs of industrial and commercial plants \cite{rahhal2020iot}. RNNs have also been deployed for data analytics in order to learn from the time series data that is typical form of the data generated by IoT and smart city deployments \cite{xie2017iot}.

\subsection{Generative Adversarial Networks (GAN)}
Generative Adversarial Networks (GAN) are mainly composed of two networks that work in conjunction to produce synthetic and high-quality data. A generative network produces data and the discriminative network distinguishes the generated data from real input data \cite{goodfellow2014generative}. The generator tries to trick the discriminator into accepting the generated data as if it were coming from a legitimate source. The two networks are thus pitted as adversaries in a GAN model.
The objective function in such a model corresponds to the existence of two networks where one tries to maximize the value function and the other tries to minimize it. If the discriminator accurately classifies the data produced by generator as being fake, then it is considered to be functioning well. Similarly, for the generator it is said to be performing well if it produces data that tricks the discriminator into accepting it as being true \cite{mohammadi2018deep}.

GANs have been used in novel pursuits to generate descriptive texts from a given image \cite{dai2017towards}, which is especially relevant for aiding visually impaired persons. They have also been used for optimizing energy consumption in IoT devices \cite{liu2018multimodal}. In addition, they have been used in healthcare domain for obtaining reliable data, which in turn support model training, and thus enable clinical decision making \cite{yang2019gan}. 

\subsection{Federated Deep Learning (FDL)}
\begin{figure}[t]
\centering
\includegraphics[width=1\textwidth, height=.4\textheight]{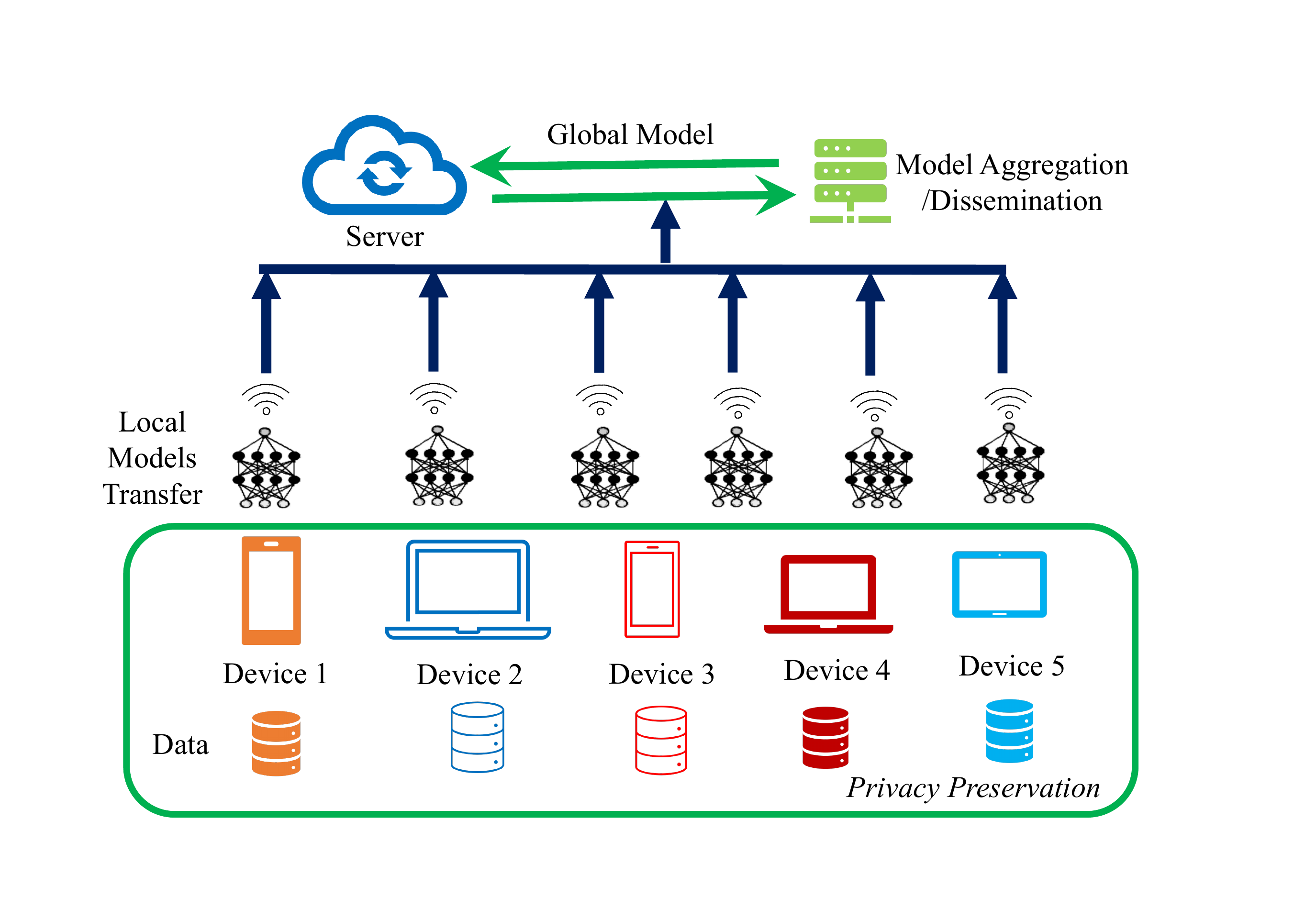}
\caption{An Overview of Federated Deep Learning}
\label{fig:FDL}
\end{figure}
A new privacy preserving collaborative paradigm is Federated Deep Learning (FDL). DL requires labeled data for training accurate models, that can perform highly structured tasks, such as classification and prediction. However, labeling a dataset has an additional cost in terms of privacy. The identification of data points can have serious implications for the generating source of such data, which are the users. Using such labeled data, it is possible to identify the source which poses a security and privacy risk. To preserve privacy in such settings, FDL has emerged as a reliable solution.

The proliferation of smartphones and the increasing popularity of mobile computing has led to the creation of application domains which are useful, but at the same time, vulnerable too. With the adoption of the smartphone, the potential for privacy violations to occur has no doubt shot up, especially since the past couple of years. A lot of the IoT devices such as wearables and medical devices collect users’ personal data and store it on users’ smartphone. Users are in general wary of sharing personal data, so is the case for other such sensitive data that may exist on the phones. Researchers have proposed that users should be able to define privacy-based policies for securing their data using novel models and mechanisms, such as attribute-based communication control (ABCC) \cite{bhatt2020abac}. However, using such data can have significant benefits in terms of getting personalized models that are trained on a variety of people. FDL provides a solution in such contexts where people may not want to share their data. By training a model locally on a user’s smartphone and then transferring just the model to a central server can help to reduce the privacy risk. Since only the model is communicated with the external server, it does not have any personal information about the user. By aggregating over a range of such local users’ model, an efficient global model can be trained which does not contain personal information. The implications of FDL are huge for mobile devices as they are the primary surface where individual model training takes place.
Figure \ref{fig:FDL} depicts how a FDL framework can deployed in IoT settings.

FDL can help to solve hitherto difficult problems in the IoT field. By using mobile devices and training learning models directly on them, user privacy is effectively preserved. Also there are benefits in terms of communication latency, cost and speed. FDL has been successfully used in Industrial IoT for enhancing quality of service by ensuring privacy preserving data sharing \cite{lu2019blockchain}. Further FDL can result in resource optimization in edge computing systems by reducing the need to share data over the network. In its place, only the model parameters can be shared. This would result in obtaining comparable model performance, which is on par with models trained using traditional ML techniques \cite{wang2019adaptive}.

%% file: System_Model.tex
\section{System Model}
\label{sec:model}

\begin{figure}[t]
\centering
\includegraphics[width=1\textwidth, height=.4\textheight]{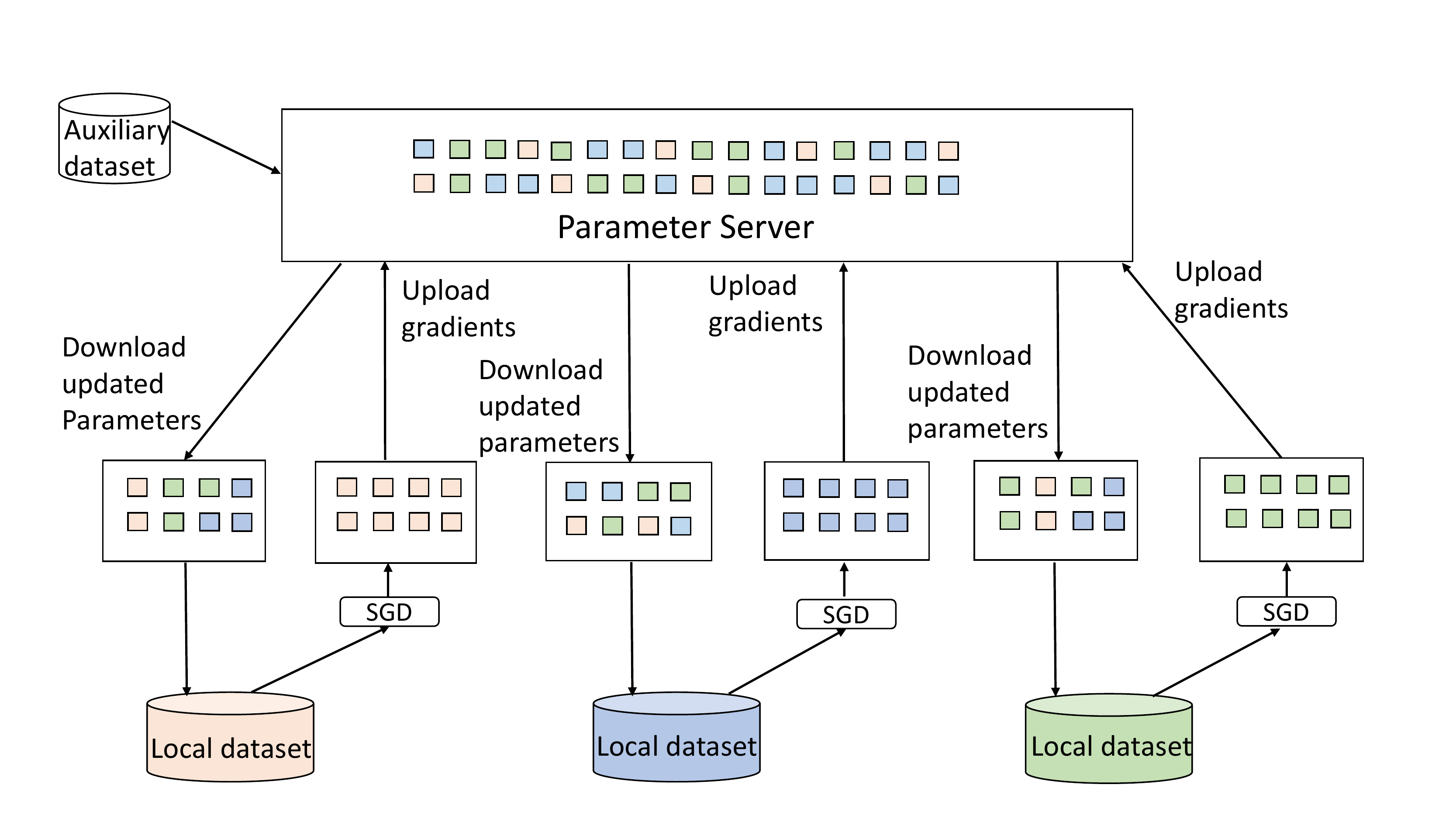}
\caption{A Collaborative Deep Learning System Model}
\label{fig:model}
\end{figure}
Here, we present comprehensive details of CDL model, where mobile edge devices or edge gateways perform training in a collaborative manner. We assume that all the mobile edge devices are altruistic in CDL model. Further, we analyze the issue of the rationality of mobile edge devices in CDL model.
\begin{figure}[t]
\centering
\includegraphics[width=0.9\textwidth, height=.5\textheight]{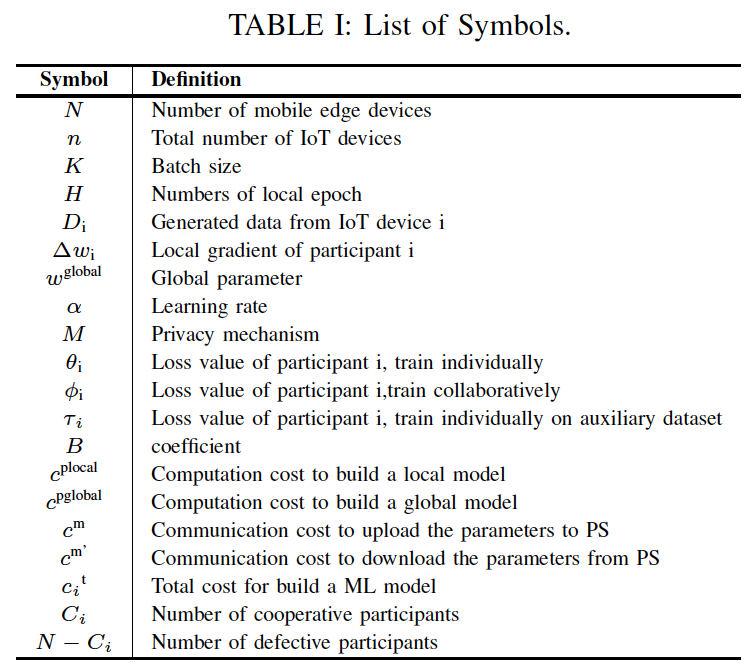}
\label{fig:table}
\end{figure}

\subsection{Collaborative Deep Learning Model in IoT}
The system model allows multiple participants to build their ML model collaboratively. Figure \ref{fig:model} presents our CDL model, which illustrates major modules of the system model. In this model, we consider that \textit{N} number of mobile edge devices or edge gateways are connected with multiple IoT devices. These IoT devices generate tremendous amount of data, which help to enhance the intelligence of their local ML model. The CDL model improves privacy of training data by exchanging local gradients instead of raw data without compromising data privacy. Each mobile edge device preserves a local vector $w^i$ of ML model, and PS also preserves another separate parameter vector $w^{global}$. After becoming a part of CDL, each edge gateway start to initialize parameters (weights) $w^i$, where \textit{i}=1,2,3,..\textit{N}, randomly. To improve the efficiency of local ML model, these initialize parameters (weights) $w^i$ are also update by downloading their updated parameters $w^{global}$ from PS.

The mobile edge devices or edge gateways participate in CDL to build their local ML model to learn a common goal. In this system model, SGD approach is used to optimize the loss value. The weight sample is selected randomly and this optimization process runs continuously until SGD reaches to a local optimum. The loss value E, which is the difference between the true value of the objective function and the computed output of the network, this value is calculated by $L^2$ norm or cross entropy. The back-propagation algorithm computes the partial derivative of E with respect to each parameter in $w^k$ and updates the parameter so as to reduce its gradient. All mobile edge devices or edge gateways build their local ML model simultaneously.

\begin{figure}[t]
\centering
\includegraphics[width=0.7\textwidth, height=.38\textheight]{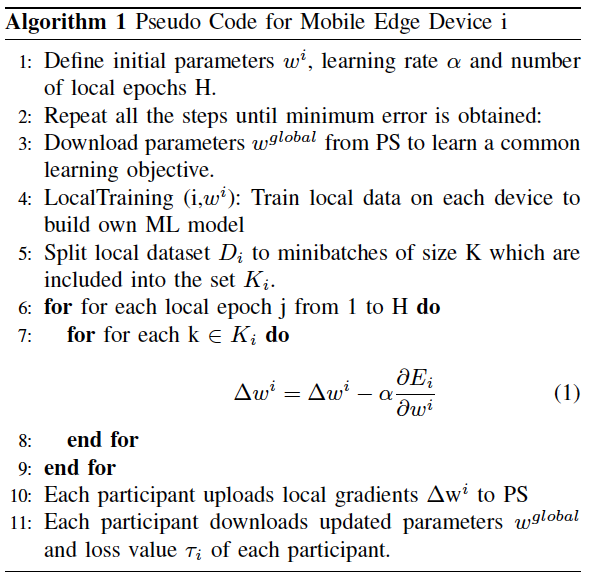}
\label{fig:algo-1}
\end{figure}

The CDL system model does not allow any one to one communication among participating mobile edge devices, however they can influence each other's training indirectly through PS. When each edge gateway receives updated parameters \textit{$\Delta$w\textsuperscript{global}} from PS. There are many ways to exchange the parameters from PS to mobile edge device. In this model, PS does not hold the process of aggregation until receiving all local gradients from all edge devices, and it works in asynchronous manner. This training and exchanging parameter process continue until the model achieves the goal.
\begin{figure}[t]
\centering
\includegraphics[width=0.7\textwidth, height=.15\textheight]{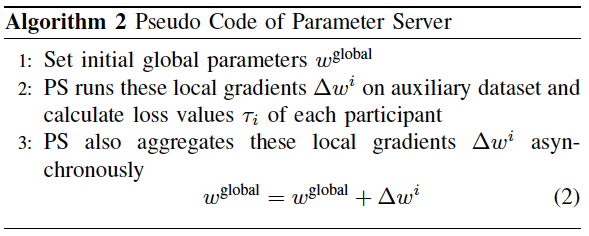}
\label{fig:algo-2}
\end{figure}

\subsection{Training Cost on Edge Gateways}
We define two major costs including computation and communication borne by mobile edge devices or edge gateways for participation in the CDL system model. There are two different phases including training and participating in this system model. Each mobile edge device or edge gateway builds a local model in the time of training and initialize their weights/parameters to train its local ML model. Then, each mobile edge device or edge gateways upload its calculated local gradients on PS in the time of participating. PS receives all the local gradients and aggregates all of them, and sends back to each mobile edge device or edge gateway. This training process continues until loss value becomes negligible.

In an epoch, the total cost of a mobile edge device to build ML model in collaborative manner based on execution of both phases. A mobile edge device pays costs \textit{c\textsuperscript{plocal}} and \textit{c\textsuperscript{pglobal}}, which are computation costs to build a local ML model and builds a local model using updated global parameters respectively in training phase. In participating phase, a mobile edge device pays another costs \textit{c\textsuperscript{m}} and \textit{c\textsuperscript{m{'}}}, which are communication costs to upload its parameters to PS and download the updated parameters from PS respectively. 
The average per mobile edge devices cost $c_{i}^{t}$ for participation in each epoch of CDL system is defined as
\begin{equation}
c_i^t = c^{plocal} +  c^{pglobal} + c^{m} + c^{m'}
\end{equation}

In a particular case, a participant may not participate in CDL, and avoids to pay some specific costs \textit{c\textsuperscript{m}}, \textit{c\textsuperscript{m'}}, and \textit{c\textsuperscript{pglobal}}. In the next section, we present rationality assumption, which provides details to avoid the pay some specific costs based on choice of strategy. 

\subsection{Rationality Assumption}
Most of research has done in distributed DL, this study \cite{chen2018machine} shows that mobile edge devices or edge gateways are controlled by an adversary. If mobile edge devices or edge gateways are behave as malicious participant, they could arbitrarily deviate from suggested protocol in CDL and they can arbitrarily drop communication between mobile edge device and PS. In our research, we assume that mobile edge devices or edge gateways and IoT devices are honest, but they behave selfish and making own benefit at minimum cost. The notion of \textit{rationality} means that a rational mobile edge device  or edge gateway decide to participate or not to participate for maximizing its profit in CDL.

%% file: Deep_Learning_Game.tex
\section{The Collaborative Deep Learning Game}
\label{sec:game}

A key contribution of this work is to show how training issue of IoT devices and mobile edge devices or edge gateways through game model. We describe the Collaborative Deep Learning game which controls the actions of multiple mobile edge devices or edge gateways and enforce them for collaboration to train their data under privacy preserving mechanism. This CDL game G is introduced with $N$-players, and these players communicate to PS simultaneously. To learn a common objective without compromising the privacy of data, each mobile edge device send local gradients to PS. PS aggregates all the gradients and sends them back, which helps to increase the accuracy of local model. However, some of edge gateways are still lacking to receive benefit from other IoT devices via gateways. 
  
\subsection{Game Theoretic Model}

Game theory is a theoretical framework for modeling conflict situations among competing players and for analyzing the behavior of various players. In the CDL game G, mobile edge devices or edge gateways, which are connected with multiple IoT devices, are participants where these gateways do not have any awareness about others and communicates to PS simultaneously. This game G is a static game, because all participants must choose their strategy simultaneously. The Game G is a tuple $(P,S,U)$, where $P$ is the set of players, $S$ is the set of strategies and $U$ is the set of payoff values. 

\begin{itemize}
\item \textbf{Players} ($P$): The set of players $P=\sum_{i=1}^{N} P_{i}$ corresponds to the set of mobile edge devices or edge gateways, where each device receives a goal from PS to build its own local model in CDL game G. 

\item \textbf{Strategy} ($S$): The CDL game G has two different strategies $S_i$ (i) Cooperative ($CP$) or (ii) Defective ($DF$), each player $P_i$ has choice between these two strategies. We refer a set of strategy as $S$ = \{$CP$, $DF$\}. These strategies determine that player $P_i$ either participates or does not participate in CDL to build the model. If a player $P_i$ chooses $CP$ strategy, it allows to send its local gradients to PS and also downloads updated parameters from PS to update its local model. There is a need to pay for various costs including communication costs ($c^m$, $c^m{}^{'}$) and computation costs ($c^{pglobal}$, $c^{plocal}$), according to $CP$ strategy. In contrast, if a player $P_i$ chooses $DF$ strategy to play, it neither uploads its local gradients to PS nor downloads the updated global parameters from PS. According to this strategy, the player pays only local computation cost $c^{plocal}$. It implies that this player is not a part of CDL and trains its local model individually on its gateway device only. 

\item \textbf{Payoff} ($U$): In CDL game G, each player's goal is to maximize their payoff, which is a function of the loss value and its various costs. Our work does not present any malicious side of players. In this game, each player receives benefit in terms of accuracy of the model and pays various costs to train the local model. 
\end{itemize}

In our CDL game G, the payoff value is depend on the loss value of the model and various costs. However, the loss value and cost value are not on the same scale. To make them similar, we introduce a coefficient B, which is multiplied by loss value. 

Now, we compute the payoff of each mobile edge device $P_i$ in this game. If we assume that the participant $P_i$ is cooperative, i.e. $P_i$ $\in$ $CP$. Similarly, if $P_i$ is defective, i.e. $P_i$ $\in$ $DF$, and the payoff $u_i$ of each mobile edge device is defined as follows.

\begin{equation}
u_i(CP) = B(\frac{1}{\phi_i}) - (c^{plocal} + c^m + c^{m'} + c^{pglobal})
\end{equation}

\begin{equation}
u_i(DF) = B(\frac{1}{\theta_i}) - (c^{plocal})  
\end{equation}
Where $\phi\textsubscript{i}$ is the loss value of the trained model using CDL and $\theta_i$ is the loss value of the trained model individually. Based on the above defined equations, we analyze our CDL game G.

\subsection{Game Analysis}
We apply the most fundamental game-theoretic concept, known as Nash Equilibrium, introduced by John Nash \cite{nash1951non} to understand the behavior of players. 

\textbf{Definition 1.} A Nash Equilibrium is a concept of game theory where none of the players can unilaterally deviate from their strategy to increase their payoff. 

In a nutshell, if both strategies present mutual best responses to each other, then no player has any motivation to deviate unilaterally from the given strategy, one Nash Equilibrium strategy profile. For instance, prisoners’ dilemma  game shows that individual players always have an incentive to choose in a way that creates a less than optimal outcome for the individuals as a group. If both players play cooperative-$CP$ strategy, it produces the best outcome for both players. In contrast, if both players decide not to cooperate with each other, they choose defective-$DF$ strategy to achieve  benefit from other players. In prisoners’ dilemma defective strategy strictly dominates the cooperation strategy. Hence, the only Nash Equilibrium in prisoners’ dilemma, is a mutual defection. 

Based on the cost and benefit of mobile edge devices to learn a neural-network model, we build a one-shot CDL game model G. In the following theorems, we show that the game G is a public good game.

\textbf{Theorem 1.} \textit{ In an iteration of collaborative deep learning game G with N mobile edge devices or edge gateways, if aggregated parameters are equally shared among all participants to build its local ML model, then this game G reduces to a public good game.} 

\textit{Proof.} We assume that all \textit{N} number of players follow defective-$DF$ strategy, and neither send their local gradients to PS nor download updated global parameters from PS. In this case, each edge gateway prefers to build local ML model individually and saves various costs including communication costs $c^m$, $c^{m'}$, and global computation cost $c^{pglobal}$. Each participant $P_i$ aims to minimize its loss value $\theta_i$ to achieves high accuracy of its ML model. None of participants cannot change his strategy profile unilaterally. Now we consider that if a participant deviates from defective-$DF$ strategy to cooperative-$CP$ strategy unilaterally, then that participant pays various costs ($c^m$ + $c^{m'}$ + $c^{pglobal}$ + $c^{plocal}$). The total payoff of cooperate-$CP$ strategy is less than defect-$DF$ strategy, so All-$DF$ is a Nash equilibrium profile and G is a public good game. 

Theorem 2 presents that we can never enforce an all cooperative-$CP$ strategy in CDL game G, and therefore, a Nash Equilibrium cannot establish when all players choose cooperative-$CP$ strategy.

\textbf{Theorem 2.}\textit{ In an iteration of collaborative deep learning game G with N mobile edge devices or edge gateways, if aggregated parameters are equally shared among all participants to build its local ML model, then we cannot establish All-Cooperation strategy profile as a Nash Equilibrium.} 

\textit{Proof.} We assume that \textit{N} number of players choose cooperative-$CP$ strategy, and ready to cooperate in collaborative deep learning. In this case, the player pays various costs including communication costs ($c^m$, $c^{m'}$) and computation costs ($c^{pglobal}$ + $c^{plocal}$). The total payoff of each player $P_i$ is calculated  by Equation (4). Now, if a player deviates from the cooperative-$CP$ to defective-$DF$ unilaterally, then that player pays only local computation cost, which is defined in Equation (5), this payoff is always greater than cooperative payoffs at Equation (4). Hence, each participant has incentive to deviate unilaterally and increases its payoff. Then, the All cooperate-$CP$ strategy profile is never a Nash Equilibrium.

\subsection{Fair Collaboration Strategy}
Each mobile edge device or edge gateway sends their local gradients to PS and these local gradients run on auxiliary dataset to calculate loss value of each dataset. Each mobile edge device downloads updated global parameters and matrix of loss values from PS. Before begin the CDL game G, each mobile edge device, which is participant has to choose its strategy to play this game G. However, in the beginning of the game, the participant is not sure about his strategy, which will depend on other participant's strategy. Therefore, all the participants are in dilemma to choose a strategy between $CP$ and $DF$. This problem is solved by proposing a novel fair collaboration strategy. K-means clustering is an unsupervised ML technique, whose purpose is to segment a data set into K clusters. Each participant applies k-means cluster algorithm on all loss values (one-dimensional data). 

\begin{figure*}[t]
\centering
\includegraphics[width=0.7\textwidth, height=.2\textheight]{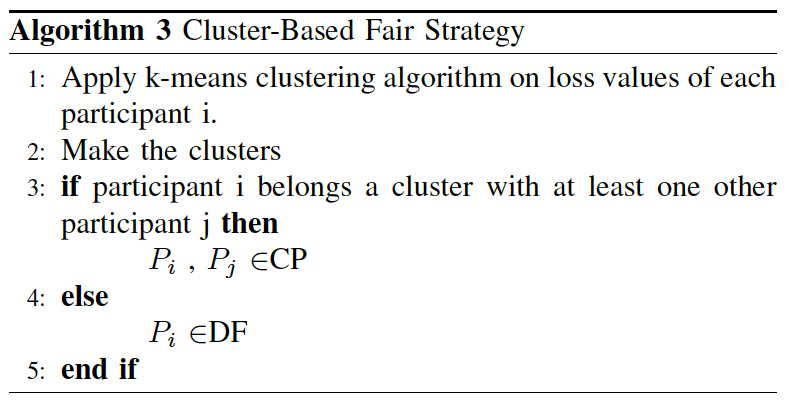}
\label{fig:algo-3}
\end{figure*}

\begin{figure}[t]
\centering
\includegraphics[width=1\textwidth, height=.4\textheight]{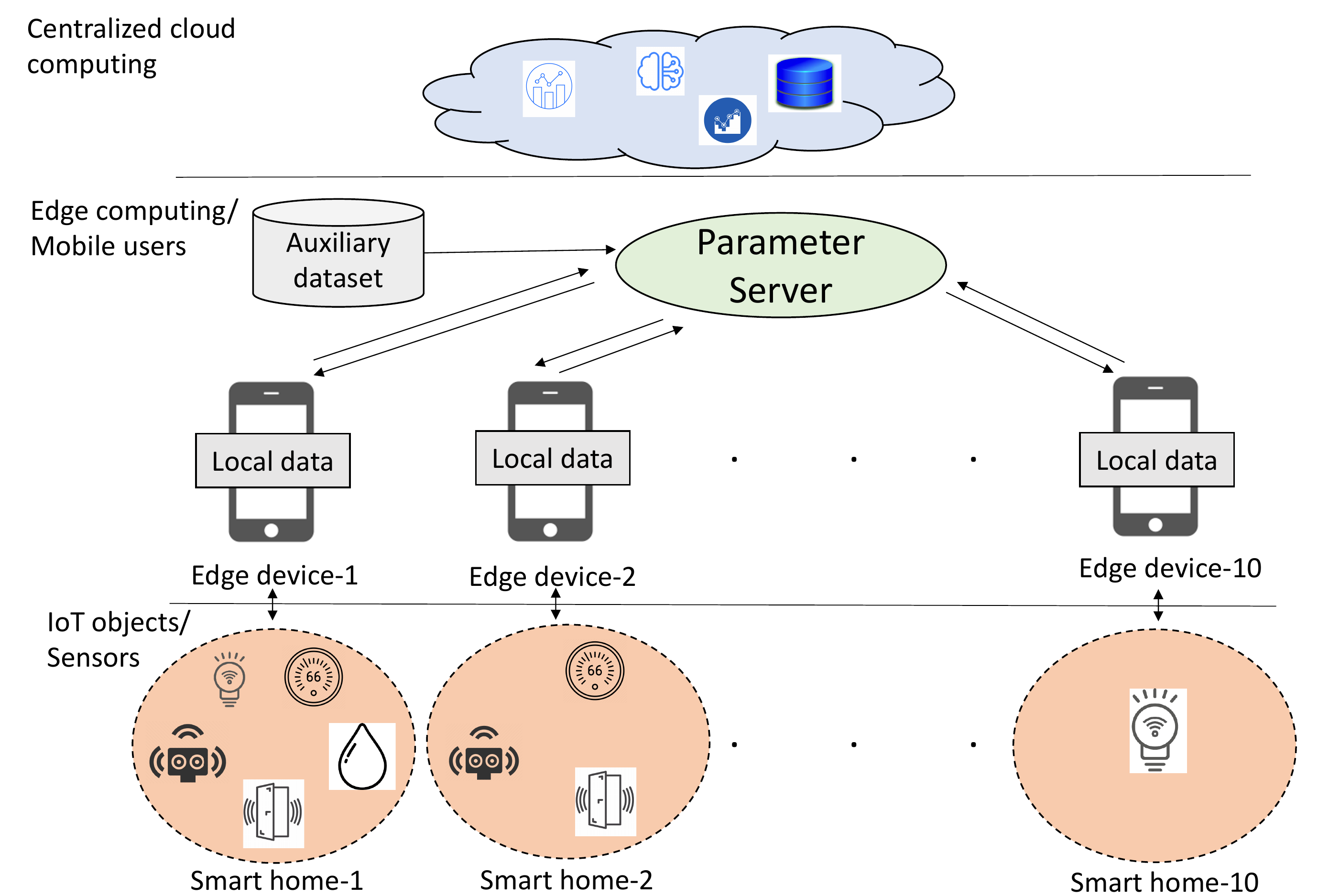}
\caption{Experimental Setup of Proposed Game Model}
\label{fig:exp}
\end{figure}

%% file: Implementation.tex
\section{Implementation and Analysis}
\label{sec:analysis}
We implemented our novel fair collaboration strategy on real-world IoT based smart home datasets for analyzing the results.  

\begin{figure*}[t]
\centering
\includegraphics[width=1\textwidth]{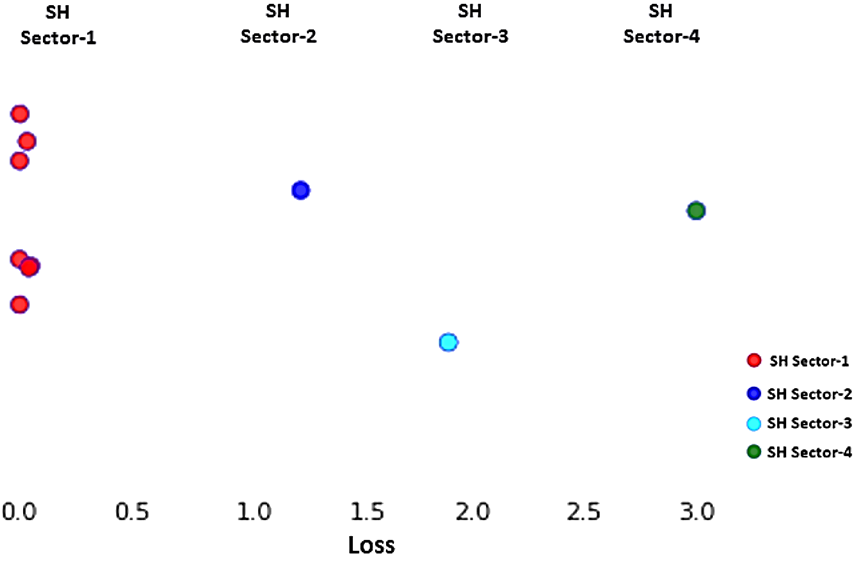}
\caption{Clustering Visualization of 10 Participants in One Dimensional Loss Value}
\label{fig:res-1}
\end{figure*}

\begin{figure}[t]
\centering
\includegraphics[width=1\textwidth]{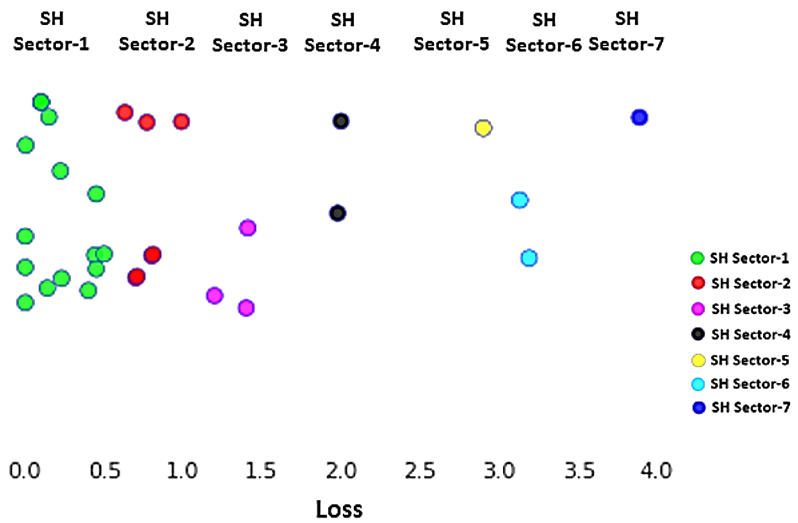}
\caption{Clustering Visualization of 30 Participants in One Dimensional Loss Value}
\label{fig:res-2}
\end{figure}

\subsection{Data Collection}
For this experiment, we select publicly available ARAS datasets \cite{alemdar2013aras} to build smart home interaction model. ARAS dataset is real-world IoT dataset, where various IoT devices setup to capture users' activities. In this smart home, the living residents did not follow any specific rules. This dataset contains two real smart home data with multiple residents for one month. It contains 3000 daily life activities captured by 26 million sensor readings in smart homes. The various sensors capture resident's various activities including washing dishes, sleeping, studying, talking on phone, and other activities. The most common sensors such as photocell, contact sensor, sonar distance, temperature sensors are attached at these smart homes. This dataset also has ground truth labels for activities, which enables to develop a new sophisticated ML smart home interaction model. 

\subsection{Data Analysis}
We present an absolute set of numerical simulations to verify our proposed fair collaborative strategy. In this experiment, we simulate the proposed system model along with mobile edge devices based on our proposed fair collaboration strategy. We performed two different experiments based on participants. First experiment of CDL is setup among 10 smart home participants, and second experiment of CDL is setup among 30 smart home participant. On the demand of our system model, we partitioned ARAS dataset unevenly into different number of participants (e.g., 10, 30 participants). Each participant (e.g., mobile edge gateway device) is connected with multiple IoT devices. Some mobile edge devices or edge gateways are connected with high number of IoT devices; however, some edge gateways are connected with low number of IoT devices. The number of IoT devices connected with edge gateway represent the quantity of dataset, while the loss value of each edge gateway represent the quality of dataset.  
For unbalanced datasets setting, the data is sorted by class and divided into two cases: (a) low quality dataset, where the participant receives data partition from a single class, and (b) high quality, where participant receives data partition from 27 classes. Figure \ref{fig:exp} shows unbalanced partitioning of the dataset, which is our first experiment. Smart home-1 generates high quality data where a high number of IoT devices are attached to the gateway device, while smart home-10 generates low quality dataset where only less number of IoT devices are attached. The following parameters are used for Algorithm 1 and 2: batch size K = 10 or 100, H = 1 or 3, $\alpha$ = 0.01. 

\subsection{Experimental Results}
In this work, we develop a novel strategy for enforcing the participants to cooperate in CDL. This proposed fair collaboration strategy brings together those participants who has similar kind of data in CDL. The game theory approach proves that participants start to leave the game if they are not getting benefit from each other. The proofs of defection is presented in section \ref{sec:game}. To validate the fair collaboration  strategy, which is based on k-means cluster, these smart-home based clusters are shown in Figure \ref{fig:res-1} and Figure \ref{fig:res-2}. The range of clusters depends on loss values of each participants. Figure \ref{fig:res-1} shows the collaboration among 10 participants using our fair collaboration strategy, and similar participants join cluster. There are 4 different clusters among 10 participants, which is referred as SH (Smart Home) Sectors. The set of each SH Sector: SH Sector-1 = \{SH1, SH2, SH4, SH5, SH7, SH8, SH9\}, SH Sector-2 = \{SH3\},SH Sector-3 = \{SH6\}, and SH Sector-4 = \{SH10\}. The above defined set indicates that smart home-1 collaborates with smart home-2, smart home-4, smart home-5, smart home-7, smart home-8, and smart home-9, while rest of the smart homes learn their local ML models individually. \\
Figure \ref{fig:res-2} also shows the cluster form of participants, and the set of each SH Sector: SH Sector-1 = \{SH1, SH2, SH3, SH4, SH5, SH6, SH7, SH8, SH9, SH11, SH17, SH18, SH19, SH20, SH26, SH27\}, SH Sector-2 = \{SH10, SH16, SH21, SH22, SH28\},SH Sector-3 = \{SH12, SH13, SH25\}, SH Sector-4 = \{SH24, SH30\}, SH Sector-5 = \{SH14\}, SH Sector-6 = \{ SH15, SH23\}, and SH Sector-7 = \{SH29\}. Some loss values for the smart homes could be same and may overlap in the evaluation figures. These graphs show that more number of participants collaborates with other participants, while individual participants are less. The overall results also show that most of participants collaborates with other participants using our proposed fair collaboration strategy in CDL. 


%% file: Conclusion.tex
\section{Conclusion and Future Work}
\label{sec:conclusion}
In this chapter, we present a system model of CDL and introduce the issue of participant rational behavior of mobile edge devices in a CDL system. We evaluate rationality of mobile edge devices in CDL using game theory model. We establish the Nash Equilibrium (NE) strategy profile for each scenario, where the learning mobile edge devices are enforced to cooperate using our novel fair collaboration strategy in CDL. 
This is a first step towards a deeper understanding of the effect of irrational participant behaviour and resulting non-cooperative behavior in CDL. As a part of the future work, we plan to propose new and revised fair collaboration strategy that can handle overlapping clusters issue by applying more efficient clustering algorithms. We also plan to calculate the accuracy of each local ML model and apply our proposed model with other IoT domain datasets. 